\documentclass{ws-procs975x65}

\begin{document}

\title{Non-singular Stiff Fluids}

\author{L. Fern\'andez-Jambrina}

\address{ETSI Navales\\
Arco de la Victoria s/n, \\ 
E-28040-Madrid, Spain\\ 
E-mail: lfernandez@etsin.upm.es}

\author{L.M. Gonz\'alez-Romero}

\address{Departamento de F\'{\i}sica Te\'orica II \\ 
Facultad de F\'{\i}sica, \\
Avenida Complutense s/n\\
E-mail: mgromero@fis.ucm.es}  


\maketitle

\abstracts{
In this talk the possibility of constructing geodesically 
complete inhomogeneous stiff fluid cosmologies is discussed. A family 
with infinite parameters is derived. A wide and easy to implement sufficient condition for 
geodesic completeness is shown.}

\section{Introduction}

The interest in geodesically complete cosmologies arises in 1990 with 
the publication of the first non-singular cosmological model by 
Senovilla\cite{seno}. Since then, the number of new regular 
inhomogeneous cosmologies has increased very little and therefore it 
has been suggested that they may be a negligible subset among 
cosmological models. 

In this talk we want to show that it is possible to obtain general 
singularity-free cosmological models depending on two nearly arbitrary 
functions\cite{wide}.

\section{Equations for $G_{2}$ stiff models}

We shall consider spacetimes with an Abelian orthogonally 
transitive $G_{2}$ group of isometries acting on timelike surfaces. 
The Killing fields are chosen to be mutually orthogonal. The metric is 
written as
\begin{equation}
ds^2=e^{2K}(-dt^2+dr^2)+e^{-2U}dz^2+\rho^2e^{2U}d\phi^2,\label{metric}
\end{equation}
\begin{equation}
    -\infty<t,z<\infty,\ 0<r<\infty,\ 0<\phi<2\pi,
\end{equation}
interpreting the isometries as cylindrical symmetry. Metric functions 
depend just on $r$ and $t$. The matter 
content is a stiff fluid, that is, a perfect fluid with pressure equal 
to energy density, $\mu=p$. 

The Einstein equations can be reduced to a simple set in convenient 
coordinates, $r=\rho$,
\begin{subequations}
\label{einstein} 
\begin{eqnarray}
0&=&U_{tt}-U_{rr}-\frac{U_{r}}{r},\label{U2}
\\
K_{t}&=&U_{t}+2r U_{t}U_{r},\label{Kr2}
\\
K_{r}&=&U_{r}+r(U_{t}^2+U_{r}^2)+\alpha r,\label{Kt2}
\\
p&=&\alpha e^{-2K},\quad \alpha>0,
\end{eqnarray}
\end{subequations}
a two-dimensional homogeneous reduced wave equation in polar coordinates
and a quadrature for $K$, which can be integrated provided we have a 
solution of the wave equation. Since the solution of the Cauchy 
problem for this equation is known\cite{john}, the problem is 
completely solved,

\begin{equation}
U(r,t)=\frac{1}{2\pi}\int_{0}^{2\pi}d\phi\int_0^1d\tau 
 \frac{\tau}{\sqrt{1-\tau^2}}\left\{tg(v)+f(v)+tf'(v) 
 \frac{t\tau^2+r\tau\cos\phi }{v}
\right\},\end{equation}
where $v=\sqrt{r^2+t^2\tau^2+2rt\tau\cos\phi}$, in terms of two 
functions $f$ and $g$. These solutions are always regular at the axis $r=0$.

\section{Geodesic Completeness}

In order to check geodesic completeness of these spacetimes we resort 
to the available theorems\cite{manolo}. The only non-trivial condition on the metric functions is,

\begin{equation}
U(0,t)\ge -\frac{1}{2}\ln |t|+b,\label{req}\end{equation} 
for a constant $b$.

\section{Polynomial Metric Functions}

A simple and wide family of functions that satisfy (\ref{req}) can 
be written in terms of polynomials. Consider

\begin{equation}
    f(r)=\sum_{i=0}^na_{i}r^i,\qquad g(r)=\sum_{i=0}^mb_{i}r^i.
\end{equation}

The leading terms in $U(0,t)$ depending on $f$ and $g$ are

\begin{equation}
    U_{f}(t)=
\sqrt{\pi}\frac{\Gamma((n+2)/2)}{\Gamma((n+1)/2)}|t|^{n},\qquad
     U_{g}(t)=
\frac{\sqrt{\pi}}{2}\frac{\Gamma((m+2)/2)}{\Gamma((m+3)/2)}|t|^{m}t.\end{equation}

There are two ways of generating a geodesically complete model:

\begin{itemize}
    \item  If $f,g$ are polynomials in $r$ respectively of degree 
    $n,m$ and $n>m+1$, we have a non-singular model if $a_{n}$  is positive.

    \item  If $f,g$ are polynomials in $r$ respectively of degree 
    $n,n-1$, $U_{f}$ and $U_{g}$ at the axis are polynomials of 
    degree $n$ and we have a non-singular model if 
    $(n+{1}/{2})\,a_{n}>|b_{n-1}|$.
    \end{itemize}
    
    Therefore, if we restrict ourselves to spacetimes with a metric 
    for which $U|_{r=0}$ is a polynomial of degree equal or lower than 
    $n$, we find that the subset of singularity-free spacetimes is an 
    open set.
    
    A simple regular model\cite{leo} may be generated for 
    $f(x)=\beta x^2/2$, $g(x)= 0$, $\beta>0$.

\section{Conclusions}

We have reduced the problem of checking the geodesic completeness of 
$G_{2}$ orthogonally transitive stiff fluid cosmologies to a simple 
condition on the behaviour of one of the metric functions at the 
axis. The case of polynomial functions has been used to generate a 
large family of singularity-free spacetimes depending on two 
functions, which are related to the initial value problem of a 
homogeneous 2D-wave equation. These results seem to point out that 
regular cosmologies cannot be considered as a negligible set.

Since these models fulfill every energy and causality 
condition\cite{HE}, the reason for their lack of singularities lies 
on the absence of trapped sets, such as closed trapped surfaces or 
compact achronous sets without edge.

Pressure is determinant for preventing the formation singularities in 
these models. From the point of view of exact solutions, it is 
interesting to mention that models obtained with a separability Ansatz\cite{agnew} are singular. This suggests a reason for the limited 
number of regular models in the literature that have been found so 
far.

\section*{Acknowledgments}
The present work has been supported by Direcci\'on General de
Ense\~nanza Superior Project PB98-0772. The authors wish to thank
 F.J. Chinea,  F. Navarro-L\'erida and  M.J. Pareja 
for valuable discussions.


\begin{thebibliography}{0}
    \bibitem{seno} J.M.M. Senovilla, \textit{ Phys. Rev. Lett.} \textbf{ 
    64}, 2219 (1990).\\
    F.J. Chinea, L. Fern\'andez-Jambrina, J.M.M. 
Senovilla, \textit{Phys. Rev. D} \textbf{45}, 481  (1992).
    
    \bibitem{wide} L. Fern\'andez-Jambrina and L.M. Gonz\'alez-Romero, \textit{Phys. Rev.} \textbf{
    D 66}, 024027 (2002) [arxiv: gr-qc/0402119].\\
    L. Fern\'andez-Jambrina, L.M. Gonz\'alez-Romero, \textit{Journ. Math.
 Phys.} \textbf{45}, 2113 (2004) [arxiv: gr-qc/0405013].

    \bibitem{john} F. John, 
    \textit{Partial Differential Equations}, 4 edition (Springer-Verlag, Berlin-New York 
    1982).

    \bibitem{manolo}L. Fern\'andez-Jambrina and L.M. Gonz\'alez-Romero, 
    \textit{Class. Quantum Grav.} \textbf{16}, 953 (1999) [arxiv: 
    gr-qc/9812039].
    \\
L. Fern\'andez-Jambrina, L.M. Gonz\'alez-Romero, \textit{Journ. Math. Phys. }
\textbf{40} {4028} (1999) [arxiv: gr-qc/9906030].

    \bibitem{leo}L. Fern\'andez-Jambrina, \textit{ Class. Quantum Grav.}, 
    \textbf{ 14}, 3407 (1997) [arxiv: gr-qc/0404017].


    \bibitem{HE} S. W. Hawking and G. F. R. Ellis, \textit{ The Large Scale 
    Structure of Space-time} (Cambridge University Press, Cambridge, 
    1973).

    \bibitem{agnew} A.F. Agnew and S.W. Goode, \textit{Class. Quantum Grav.} 
    \textbf{11}, 1725 (1994).

\end{thebibliography}
\end{document}